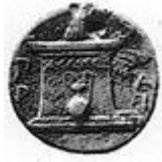

# HAROKOPION UNIVERSITY OF ATHENS
## DEPARTMENT OF INFORMATICS & TELEMATICS

**Master Thesis Title:**

# IMPLEMENTATION OF FUNCTIONS IN R TOOL IN PARALLEL ENVIRONMENT

**Supervisor**
*Konstantinos Tserpes, Lecturer*
**Supervisory Committee**
*Iraklis Varlamis, Assistant Professor*
*Dimitrios Michail, Assistant Professor*

**Antonios Makris(Id.N.: 13108)**
*MSc candidate in Web Engineering*
*BSc in Informatics and Telematics*

Athens, July 2015

# Acknowledgements

*As the author of this thesis, i would like to express my sincere appreciation to all those who assisted and supported me in the preparation of this master thesis as well as throughout my studies in general.*

*I would like to thank the supervisor of my thesis and Lecturer at the Department of Informatics & Telematics in Harokopio University Kostantino Tserpe , the Assistant Professor Irakli Varlami and Assistant Professor Dimitrio Michail, for their assistance and substantial help in the completion of my thesis, the excellent collaboration, the valuable guidance and knowledge and for the strong support and confidence. Also, I would like to thank them for the opportunity they gave me to work on research issues and expand my horizons.*

*Finally, I would like to thank my family and friends for their support throughout my studies and for their understanding and patience.*

# Abstract

Drug promiscuity and polypharmacology are much discussed topics in pharmaceutical research. Drug repositioning applies established drugs to new disease indications with increasing success. As polypharmacology, defined a drug's ability to bind to several targets but due to possible side effects, this feature is not taken into consideration. Thus, the pharmaceutical industry focused on the development of highly selective single-target drugs. Nowadays after lot of researches, it is clear that polypharmacology is important for the efficacy of drugs. There are side effects but on the other hand, this gives the opportunity to uncover new uses for already known drugs and especially for complex diseases. Thus, it is clear that there are two sides of the same coin.

There are several approaches to discover new drugs targets, as analysis of genome wide association, gene expression data and networks, structural approaches with alignment methods etc. Computational drug discovery and design has experienced a rapid increase in development which is mainly due to the increasing cost for discovering new compounds. Since drug development is a very costly venture, the pharmaceutical industry puts effort in the repositioning of withdrawn or already approved drugs. The costs for bringing such a drug to market are 60% lower than the development of a novel drug, which costs roughly one billion US dollars. Thus, target prediction, drug repositioning approaches, protein-ligand docking and scoring algorithms, virtual screening and other computational techniques have gained the interest of researchers and pharmaceutical companies.

According to researches, drug repositioning process makes use of technologies and algorithms based on sequential programming. Thus it presents research interest the parallelization of such procedures, in order to make improvements in the accuracy and performance of the system in comparison to the current art.

**Keywords**: Drug promiscuity, Polypharmacology, Binding site, Structural Alignment, Protein Structures, Parallelization, Isomorphism, Sequence Alignment

# Abstract in Greek

Η "ασυδοσία" των φαρμάκων και η πολυφαρμακολογία, αποτελούν πολυσυζητημένα θέματα στη φαρμακευτική έρευνα. Η επανατοποθέτηση φαρμάκων εφαρμόζεται στα ήδη υπάρχοντα φάρμακα για την ένδειξη ασθενειών, με αυξανόμενη επιτυχία. Ως πολυφαρμακολογία, ορίζεται η ικανότητα ενός φαρμάκου να συνδέεται με διάφορους στόχους αλλά λόγω πιθανών παρενεργειών, αυτό το χαρακτηριστικό δεν είχε ληφθεί σοβαρά υπόψη. Έτσι, η φαρμακευτική βιομηχανία επικεντρώθηκε στην ανάπτυξη φαρμάκων μοναδικού στόχου. Σήμερα μετά από έρευνες, είναι σαφές ότι η πολυφαρμακολογία είναι σημαντική για την αποτελεσματικότητα των φαρμάκων. Υπάρχουν παρενέργειες, αλλά από την άλλη πλευρά, δίνεται τη δυνατότητα ανακάλυψης νέων χρήσεων για ήδη γνωστά φάρμακα και ειδικά για πολύπλοκες ασθένειες. Έτσι είναι σαφές, ότι υπάρχουν δύο όψεις του ίδιου νομίσματος.

Υπάρχουν διάφορες προσεγγίσεις για την ανακάλυψη νέων στόχων όσο αφορά τα φάρμακα, όπως η ανάλυση του γονιδιώματος σε επίπεδο σύνδεσης, γονιδιακά δεδομένα έκφρασης και δίκτυα, δομικές προσεγγίσεις με μεθόδους ευθυγράμμισης κτλ. Η υπολογιστική ανακάλυψη φαρμάκων και ο σχεδιασμός αυτών, γνώρισε ραγδαία ανάπτυξη, η οποία οφείλεται κυρίως στο αυξημένο κόστος, για την ανακάλυψη νέων ενώσεων. Δεδομένου ότι η ανάπτυξη φαρμάκων είναι μια πολύ δαπανηρή διαδικασία, η φαρμακευτική βιομηχανία έχει επικεντρωθεί στην επανατοποθέτηση αποσυρθέντων ή ήδη εγκεκριμένων φαρμάκων. Οι δαπάνες για την δημιουργία και διοχέτευση ενός τέτοιου φαρμάκου στην αγορά είναι 60% χαμηλότερες από την ανάπτυξη ενός νέου, το οποίο κοστίζει περίπου ένα δισεκατομμύριο δολάρια. Έτσι προσεγγίσεις που αφορούν, την επανατοποθέτηση φαρμάκων, την πρόβλεψη πρωτεϊνικών στόχων, τους αλγόριθμους σύνδεσης, την βαθμολόγηση πρωτεΐνης-προσδέτη και άλλες υπολογιστικές τεχνικές έχουν κερδίσει το ενδιαφέρον των ερευνητών και των φαρμακευτικών εταιρειών.

Σύμφωνα με έρευνες, η διαδικασία επανατοποθέτησης φαρμάκων επιτυγχάνεται μέσω τεχνολογιών και αλγορίθμων, που βασίζονται στον σειριακό προγραμματισμό. Έτσι παρουσιάζει ερευνητικό ενδιαφέρον, η παραλληλοποίηση αυτών των διαδικασιών με σκοπό την βελτίωση της ακρίβειας και της απόδοσης του συστήματος, σε σύγκριση με την τρέχουσα τεχνολογία.

# Table of Contents

# 1. Introduction - The problem

Drug discovery and development is a complex and expensive process. Due to the exponential growth of protein and molecular data and fast advancement in technologies, the efforts of drug discovery have been tremendously amplified.

Early efforts in drug discovery relied on screening natural products derived from plants and microorganisms and testing them for activity in animal models. This was a slow and labour-intensive process carried out mostly in the wet lab.

Computational drug discovery and design has experienced a rapid increase in development which is mainly due to the increasing cost for discovering new compounds. According to researches the cost for producing a new drug is estimated to have grown at an annual compound rate of 13.4% since the 1950s. The cost for developing a new drug is on average 4 billion US$ and requires 15 years until it reaches the market. The clinical trials are the most expensive stages in this process. However, the research for a target that links to a disease and a molecule that modulates the function of target without side effects mostly influences approval at early stages and increases the clinical target validation success rate. As a result target prediction, drug repositioning approaches, protein-ligand docking and scoring algorithms, virtual screening and other computational techniques have gained the interest of researchers and pharmaceutical companies.

The most recent and successful research on drugs repositioning, makes use of technologies and algorithms based on sequential programming.

Generally in this study, we investigate technologies which tackle the problem of the identification of structurally similar proteins in order to identify new purposes for the use of drugs. The current study, aims at identifying combinations of proteins that may have similarities with a view to repositioning them. We implement a solution based on Open Message Passing Interface (OpenMPI) with the purpose to make improvements in the accuracy and performance of the system in comparison to the current art.

The main plan is to conduct experiments based on the pipeline algorithms proposed by Haupt et al (2013). These pipeline algorithms were executed sequentially and for specific inputs. Our goal is to parallelize the whole process but also part of this in order to improve the execution speed, avoid bottlenecks and extract quickly results.

The key-concept of the advances that this work brings is based on the performance of simulations of computational drug-target predictions. Drug-target binding simulations aim at the advancement of medical and biological knowledge at a fast pace and a low cost.

**1.1 Structure**

Section 2 presents the current state of the art system and technologies for the Binding Site Similarity process.

Section 3 presents the design of our system in general, the technologies used, the packages and libraries that installed and the architecture of the system.

Section 4 presents the implementation conducted, the functions that used and the parallelization of both levels of the BSS pipeline.

Section 5 presents the results of experiments. In general, compared and represented the execution time among different setups and levels of parallelization.

Section 6 presents the conclusion of this study and the future work plans.

# 2. Related Work

## 2.1 Introduction - Drug Promiscuity

Drug repositioning applies established drugs to new disease indications with increasing success. A prerequisite for drug repurposing is drug promiscuity (polypharmacology) ie a drug's ability to bind to several targets. Drug promiscuity remains an open issue and hydrophobicity and molecular weight have been suggested as key reasons. Protein structures offer a structural dimension to explain promiscuity and here raises the question if a drug bind multiple targets because it is flexible or because the targets are structurally similar or share similar binding sites. [11]

Developing a drug in laboratory is a laborious and costly endeavor. Thus, the repositioning of already approved drugs for the treatment of new diseases is promising and valuable. One computational approach to repositioning exploits the structural similarity of binding sites of known and new targets and also reduces costs and improves efficiency. So, the pharmaceutical industry puts effort in the repositioning of withdrawn or already approved drugs. The costs for bringing such a drug to market are 60% lower than the development of a novel drug, which costs roughly one billion US dollars. Drug repositioning is the process of finding new indications for existing drugs. Beside the lower drug development costs and the reduced time for approval, a new drug application can help to expand the patent life and increase the return for the investment in the development of the drug. [12]

Binding sites are closely related to protein function. The identification of binding sites in proteins is essential to the understanding of their interactions with ligands, including other proteins. Binding sites can retain conservation of sequence and structure. Structural conservation however is more prevalent. Even in the absence of obvious sequence similarity, structural similarity between two protein structures can imply common ancestry, which in turn can suggest a similar function. [13]

Drug repurposing is supported by the core observation that a single drug often interacts with multiple targets. It offers an appealing strategy and enables reuse of existing therapeutic compounds. Predicting off-targets by computational methods is gaining increasing interest in early-stage drug discovery. When a similar binding site is detected in the Protein Data Bank (PDB), it is possible that the corresponding ligand also binds to that similar site. This target hopping case is probably rare because it requires a high similarity between the binding sites and on the other hand, it could be a strong rationale evidence to highlight possible off-target reactions and possibly a potential undesired side effect. This target-based drug repurposing can be extended a significant step further with the capability of searching the full surface of all proteins in the PDB, and therefore not relying on pocket detection. Resolved 3D protein structures are a major source of information for understanding protein functional properties. The current explosive growth of publicly available protein structures in the Protein Data Bank (PDB) is producing massive volumes of data for computational modeling and drug design methods. [14]

The prerequisite for drug repositioning is polypharmacology (drug promiscuity), meaning that one drug binds to multiple distinct targets. Recent researches have shown that promiscuity of proteins in ligand binding as well as in function is not as rare as previously thought. There is an enormous potential to find novel targets for already known drugs based on the approved targets since promiscuous drugs are common. If a drug target defined as a protein to which a drug molecule binds physically, there are about 320 known targets for approved drugs. By applying a more loose definition and including experimental drugs can easily lead to 6000 or more drug targets. A study of a data set of 276.122 bioactive compounds revealed that 35% of them are known to bind to more than one target. A quarter of these was found to bind to proteins from different gene families.

The permissive binding of a drug to off-targets can be the cause of adverse side effects but may in contrast also increase its efficacy. Similarities in the key pharmacophores of structurally different proteins can lead to high affinity binding of the same drug. According to researches, one way to identify promiscuous proteins is chemocentric, which exploits chemical similarity among the ligands but this approach is weakened because chemical similarity seems to contribute much less to similarity in the biological activity profile. Another method is the binding site centric chemical space. To find new targets leading to repositioning drugs there are the following techniques: a) Non-in silico, b) In vitro, c) In silico and should apply a combination thereof, in order to extract correct results. [12]

## 2.2 Binding Site Similarity (BSS)

Earlier, a drug binding to multiple different targets was unwanted in drug development due to possible side effects. For this reason, the pharmaceutical industry focused on the development of highly selective single-target drugs. Nowadays, it is clear that polypharmacology (drug promiscuity), is important for the efficacy of drugs. This gives the opportunity to uncover new uses for already known drugs and increase the efficacy of drugs. There are efforts to develop promiscuous drugs, especially for complex diseases. Approaches to discover new drug targets and uses are manifold, ranging from the analysis of genome wide association studies, gene expression data and networks to structural approaches. Structural binding site comparison approaches can be distinguished in alignment methods and alignment-free methods. These methods have the advantage of uncovering also distant similarities but do not provide an aligned structure. [11]

According to researches, there is no correlation between drug promiscuity and molecular weight or hydrophobicity. A drug may be promiscuous for reasons of structural nature. This means that a drug may be flexible and can bind to different binding sites or the binding sites of the targets may be similar as shown in Figure 2.1 below.

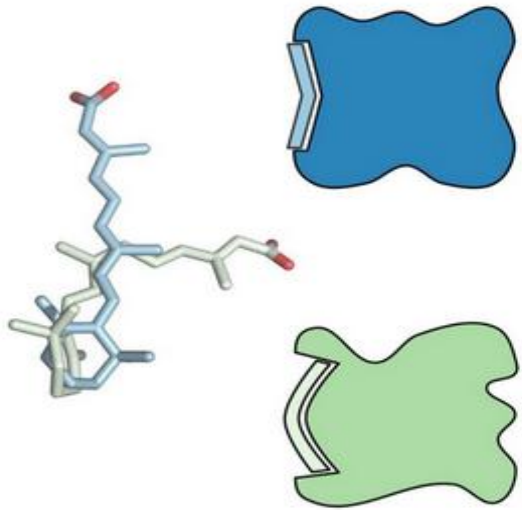

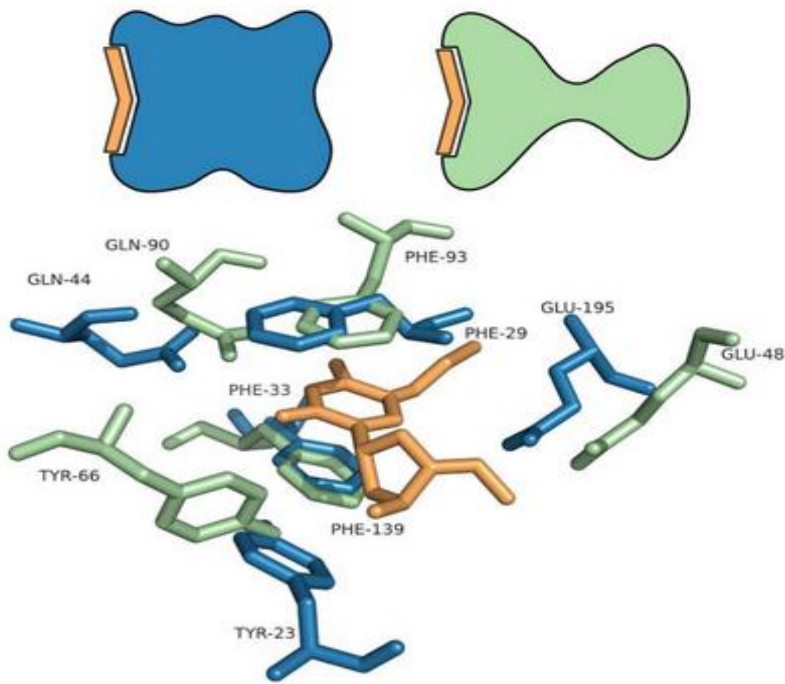

Figure 2.1 Drug promiscuity: Ligand flexibility vs binding site similarity (Haupt et al., 2013)

For the needs of this research, there were selected 543 drugs from PDB and 164 of those with three or more targets. These drugs bind a set of 712 non-redundant protein structures and clustered by 95% sequence identity.

To investigate whether a protein's pair has a similar binding site, SMAP align locally their structures. Initially, it made selection of the Ca atoms from each protein and then found the local superposition of these atoms in space. Ligand positions were used to evaluate alignment, since only binding sites of identical promiscuous drugs are aligned against each other. Ligand positions can be found, by measuring distances between the atoms of the ligands (RMSD) that we examine. Subsequently, it conducted an automated substructure search using SMSD (Small Molecule Subgraph Detector), generating a robust scoring for binding site similarity: LigandRMSD. The pipeline that shows whether promiscuous drugs have targets with similar binding sites affected in these steps:

1. Alignment of all pairs of binding sites for all promiscuous drugs using SMAP
2. Removing redundant targets
3. Keeping sites with a consistent binding mode of the ligand

In step 1 it conducted comparison of any possible match of binding sites in a promiscuous drug. Since the comparison is pair wise, the number of 712 non-redundant targets (2284 structures) leads to a total 38244 aligned binding site pairs.

In step 2 redundant targets removed by clustering targets with 95% sequence identity. This is important because some of the aligned binding sites are very similar and thus it reduced the dataset to nearly 10% (3948 pairs)

In step 3 it conducted comparisons between the ligands, since it is vital that the compared binding sites, bind the ligand in a similar mode. This step is called LigandRMSD and made a computation of how well the two ligands of the compared binding sites are aligned due to the superposition of the binding sites, by measuring the RMSD of the ligand superposition and comparing it to a corresponding optimal superposition. The alignment of two binding sites is successful if the positions of their shared ligand in each site are similar. Thus this step reduces the set by 59% and final set consists 1628 binding site pairs. The pipeline of the binding site similarity analysis is shown in Figure 2.2.

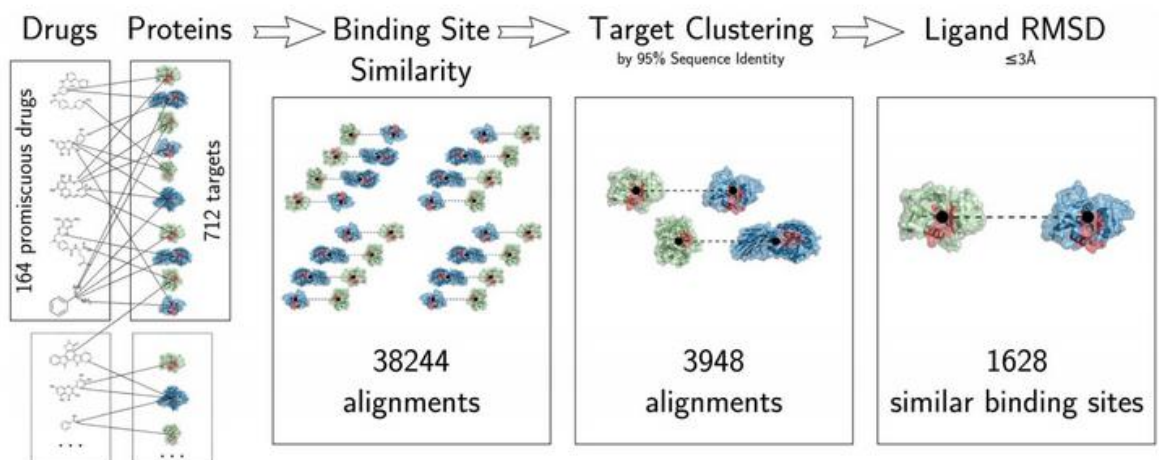

Figure 2.2: Pipeline of the binding site similarity analysis (Haupt et al., 2013)

1628 out of the 3948 target pairs (41%) have a similar binding site according to the alignment by SMAP together with the scoring with LigandRMSD. The average sequence identity of these 1628 target pairs is still low with 28% and the majority (1112 pairs) has less than 30% sequence identity. Taking a drug-centric view, 71% of the drugs have at least one target pair with a similar binding site and that for 18% of the drugs all of their targets are similar. Conclusively, according to this research similar binding sites and structural similarity correlate with promiscuity. [11]

There have been a number of approaches to computational drug repositioning, including similarity of side effects, similarity of gene expression profiles of different diseases and structural similarity of binding sites. Structural similarity of binding sites provides insight into the mode of action of the drug. Many researches of proteins have shown, that similar binding sites exist among different proteins and similar binding sites most likely bind to the same ligands. This observation can be exploited, using binding site comparison methods to find new targets for known drugs leading eventually to drug repositioning as shown in Figure 2.3. The examination of protein–drug interactions helps to understand drug modes of action and reduce drug doses. The identification of off-targets gives the opportunity to optimize drugs to gain a higher selectivity and thus reduce side effects.

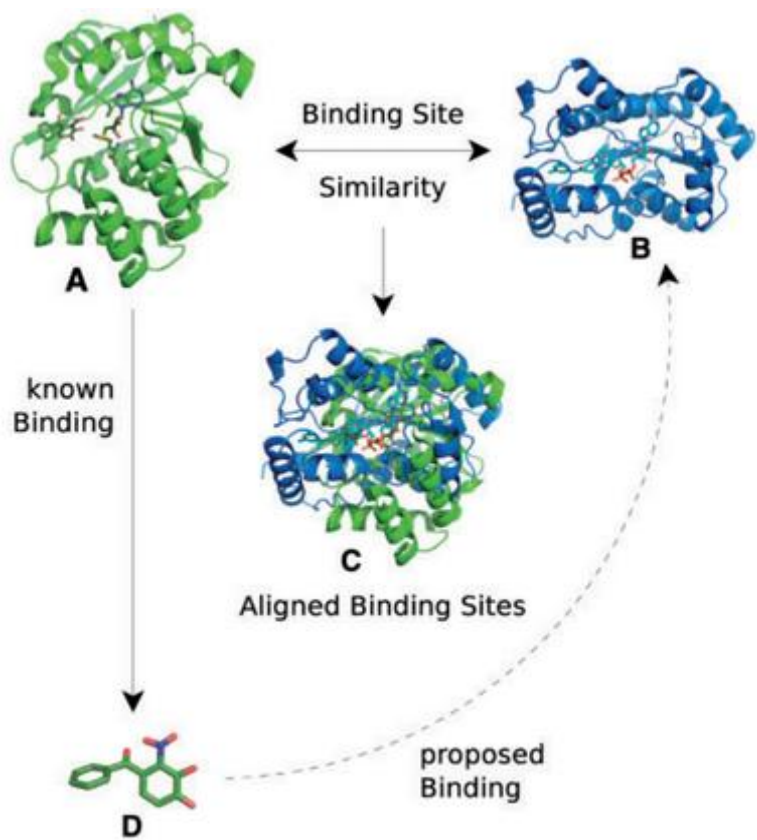

Figure 2.3: Drug repositioning (Haupt and Schroeder, 2011)

The figure above represents the drug repositioning process using binding site similarity. Two proteins A and B have similar binding sites and thus can be aligned (C). The found binding site similarity suggests that the ligand (D) may also bind to the protein (B). This gives a candidate for drug repositioning.

One of the reasons for drug promiscuity constitutes the similarity of binding sites among proteins. These proteins either belong to the same family, thus showing a similar tertiary structure, or are distinct in their overall structure. The binding site centric chemical space seems to become the new technique in structure-based drug discovery and together with the ever growing structural data in the PDB, gives the opportunity to extensively use protein structural data and thus to find new drug targets and candidate drugs for repositioning. This method aims to reveal structural similarity in protein binding site pairs, although not necessarily similar in sequence or fold. Searching the whole PDB for a specific binding site can identify unexpected potential off-targets right at the beginning of the drug discovery process. [12]

## 2.3 Alignment Algorithms for Binding Site Similarity

The current algorithms for binding site comparison tools, allow the quantification of local similarity between protein structures. These algorithms are based mostly on geometrical comparisons. Basic thing in them is that they abstract from the binding site, converting it into a discrete simplified representation. Binding sites representation can be accomplished in many different ways such as pharmacophoric features, atoms or residues. These are represented by one or more points and are labeled according to their physicochemical or geometric properties and only surface residues are considered. The binding pockets of a protein are identified by the analysis of the surface. Thus, there are also some algorithms that analyze the complete protein surface. The points can be assigned grid-based or irregularly, represent parts of the residues, individual atoms or neighborhood properties. Such points are called features. The resulting feature patterns of two binding sites are then structurally aligned such that the number of matched features is optimized. The binding site features are either represented as geometric patterns or as numerical fingerprints.

First of all, the local structural alignment algorithms create the reduced geometrical representation of the binding site. Thus, the complexity of the pair-wise comparison is reduced. After that, properties are assigned to the points and finally, the binding sites are aligned. The approaches for binding site alignment are the following:

- Iterative search for the best translation/rotation. The largest alignment is considered best
- Geometric matching
- Geometric hashing, which maps all triangles similar in shape, regardless of their orientation, to identical or very close hash slots
- Clique detection
- SiteAlign, that based on the binding sites numerical fingerprints
- Alignment-free binding site comparison [12]

**2.3.1 Scoring Functions and Success Alignment**

The binding site alignment is evaluated through a scoring function. The scoring can be directly incorporated in the alignment process, excluding bad scoring alignments. A key influence to the score is the number of aligned points. Basic scores are:

- Tanimoto-index (number of aligned features divided by the total number of features)
- RMSD for geometry or other properties
- Amino acid substitution (identification of similar binding sites)
- Normalization scores permit the ranking of different alignments of one query to different target binding sites. A possible normalization is dividing the target–query score by the query–query score.

There are also three parameters which influence the success of the site alignment and are the following: a) resolution of the binding site, b) distance tolerance and c) the complexity of feature descriptors. Methods, which use a comprehensive representation, are more prone to false negatives, which is due to slight changes in the binding sites in comparison to Ca atom-based descriptions. [12]

**2.4 Similarity between 3D protein structures - Comparison of protein binding sites**

Quantification of local similarity between protein 3D structures is a promising tool in computer-aided drug design and prediction of biological function. Almost all computational methods are based on geometrical comparisons. Structure-based methods for local comparison of unrelated proteins typically use a simplified representation of cavity residues, by partitioning of the surface into small patches which are then treated as geometric patterns or numerical fingerprints.

All the described methods are based on geometric pattern comparisons that follow the same three steps (Figure 2.4): a) the structures of the two proteins are parsed into meaningful 3D coordinates in order to reduce the complexity of the pairwise comparison, b) the two resulting patterns are structurally aligned using the transformation that produces the maximum number of equivalent points and c) a scoring function quantifies the similarity based on aligned features.

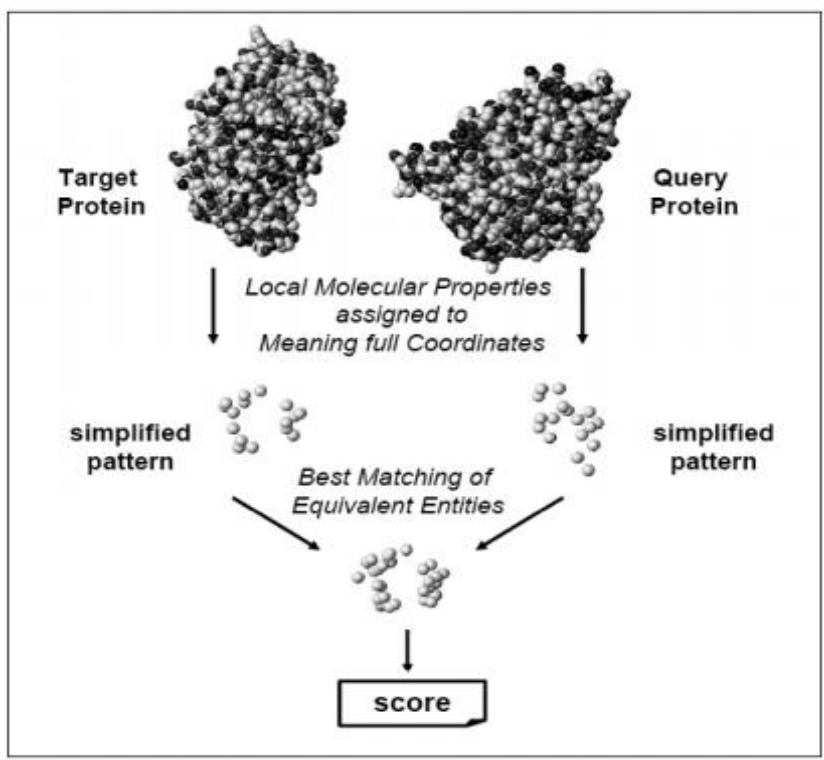

Figure 2.4: Overview of the comparison of protein binding sites (Kellenberger et al., 2008)

The methods that are based on geometric pattern comparisons are divided into two categories:

1. Simplified Representation of Protein Cavities. The basic assumption in the search for similarity between active sites is that only residues close to the surface need to be considered. These amino acids may directly or indirectly be involved in ligand recognition. All surface residues or only cavity-flanking residues are taken into account. The cavity is either deduced from the analysis of the protein surface shape or defined from the distance of cavity-lining residues to a co-crystallized ligand. Whereas the former method only selects solvent accessible surface residues, the latter method may include a few buried residues. Cavity residues are then transformed into either an irregular or a regular 3D arrangement of points.

2. Geometric Search for the Best Structural Alignment. The similarity between the cavities of two proteins is always inferred from the best structural alignment of corresponding patterns.

a. iterative search for the best translation/rotation
b. geometric matching
c. geometric hashing (Figure 2.5 A)
d. clique detection (Figure 2.5 B) [15]

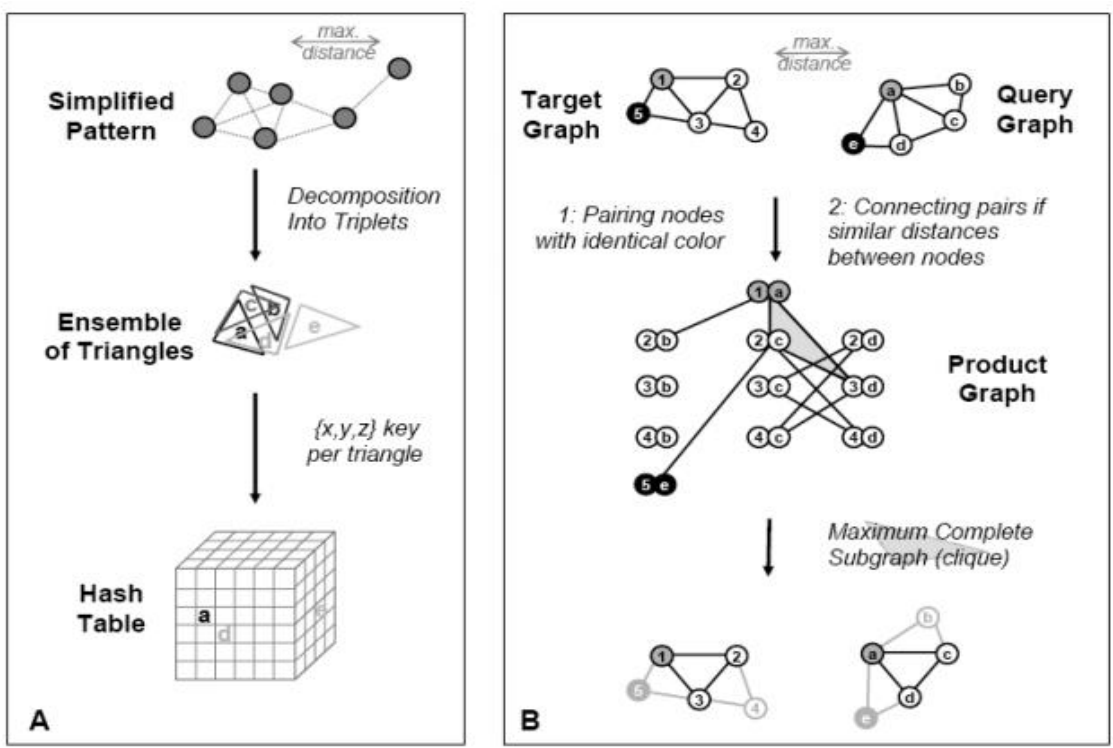

Figure 2.5 (A) Geometric hashing. Because of the inter-point distance threshold used to generate the triplets, no triangles include the far-right point. For the sake of clarity, no labels were associated to points. (B) Clique detection. In the graph definition, nodes are connected if closer than a distance threshold. Point's properties are represented by different achromatic colors. (Kellenberger et al., 2008)

## 2.5 Detection of structurally similar protein binding sites by local structural alignment

Binding sites of protein structures are closely related to protein function. The identification of binding sites in proteins is essential to the understanding of their interactions with ligands and other proteins. Binding sites can retain conservation of sequence and structure. Even in the absence of obvious sequence similarity, structural similarity between two protein structures can imply common ancestry and a similar function, but it is also possible for structurally similar proteins to have different functions. Similar folding by itself does not necessarily imply evolutionary divergence. In the case of divergent evolution, similarity is due to the common origin, such as accumulation of differences from homologous ancestral protein structures. The frequency of two different folds converging to a similar statistically significant side-chain pattern is ~1%. If a protein has a known 3D structure but no known function, inferences concerning function can be made by comparison to other proteins.

The algorithm used for exploitation of locally similar 3D patterns of physicochemical properties on the surface of a protein for detection of binding sites that may lack sequence and global structural conservation, is ProBis.

### 2.5.1 ProBis Algorithm

ProBiS can identify protein–protein binding sites by searching for conserved protein surface structure and physicochemical properties in proteins with similar folds and permitting comparison of only a few structural neighbors. The method was conducted in following steps (Figure 2.6):

A. A query protein structure (Q) whose structurally similar regions are to be detected is compared with each of ~23 000 non-redundant PDB structures (P). Residues on the surface of the proteins are represented as graphs with vertices and edges. Vertices that are separated by <15 Å are connected with edges.

B. Proteins are represented as graphs of vertices and edges, are divided into $n$ overlapping subgraphs, where $n$ is the number of vertices in each protein graph. To find if two subgraphs represent similar parts of their respective protein surfaces, the two distance matrices are subtracted, and from the resulting difference matrix, a value of similarity is calculated. For each pair of query and database protein's subgraphs conducted a graph.

C. All sufficiently similar pairs of subgraphs that pass the filtering in the previous step, subjected to the more rigorous maximum clique procedure, which detects vertex-to-vertex correspondences between the two protein graphs being compared. Then, the algorithm finds a maximum clique in each product graph.

D. Each maximum clique is equivalent to a single structural alignment and superimposition of two compared proteins. The alignments are local, and allow superimposition of maximum numbers of two protein subgraph vertices.

E. Steps A–D are repeated for each database protein and the resulting alignments and their scores are saved to a results database implemented on the MySQL platform.

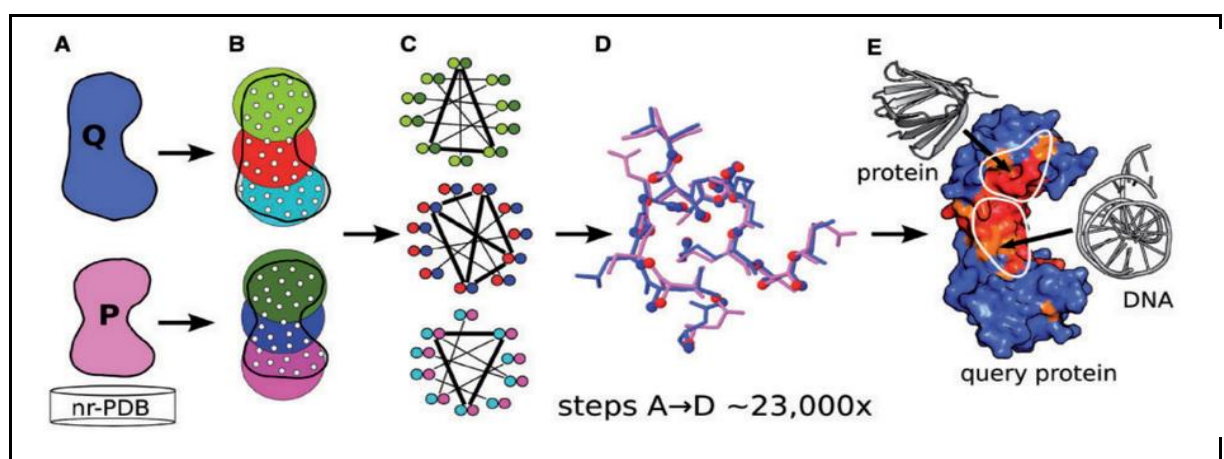

Figure 2.6: ProBis Algorithm (Konc and Janezic,2010)

## 2.5.1.1 Scores for structural alignment which performed by ProBis Algorithm

There are five distinct criteria that are applied to each structural alignment. They are used to measure the statistical and structural significance and to permit filtering out off insignificant alignments. Local structural alignments do not need to involve a large number of residues, thus the similarity that is detected could in fact be a frequently occurring 3D motif, having no relationship to common protein function.

Score measures:
- Surface vectors angle: for each of the two superimposed sets of vertices, an outer-pointing surface vector originating in the geometric center and constructed perpendicular to the surface of the protein
- Surface patch RMSD: measures the shape similarity of the two superimposed surfaces patches

- Surface patch size: discard each structural alignment with fewer than ten vertices
- E-value [13]

**2.6 Methods for ligand binding**

There are several and different computational methods for the detection and characterization of protein ligand-binding sites. These methods, are divided in the following areas: *functional site detection*, whereby protein evolutionary information has been used to locate binding sites on the protein surface, *functional site similarity*, whereby structural similarity and three dimensional templates can be used to compare and classify and potentially locate new binding sites and *ligand docking*, which is being used to find and validate functional sites. [16]

# 3. Design

## 3.1 System architecture

Initially, our infrastructure takes as input a protein pair. Subsequently, we installed R language and SPRINT framework in each node. Also in each node we installed jdk, mpi, python, openbabel, pybel, rdkit and eigen. The illustration below (Figure 3.1) represents the system architecture:

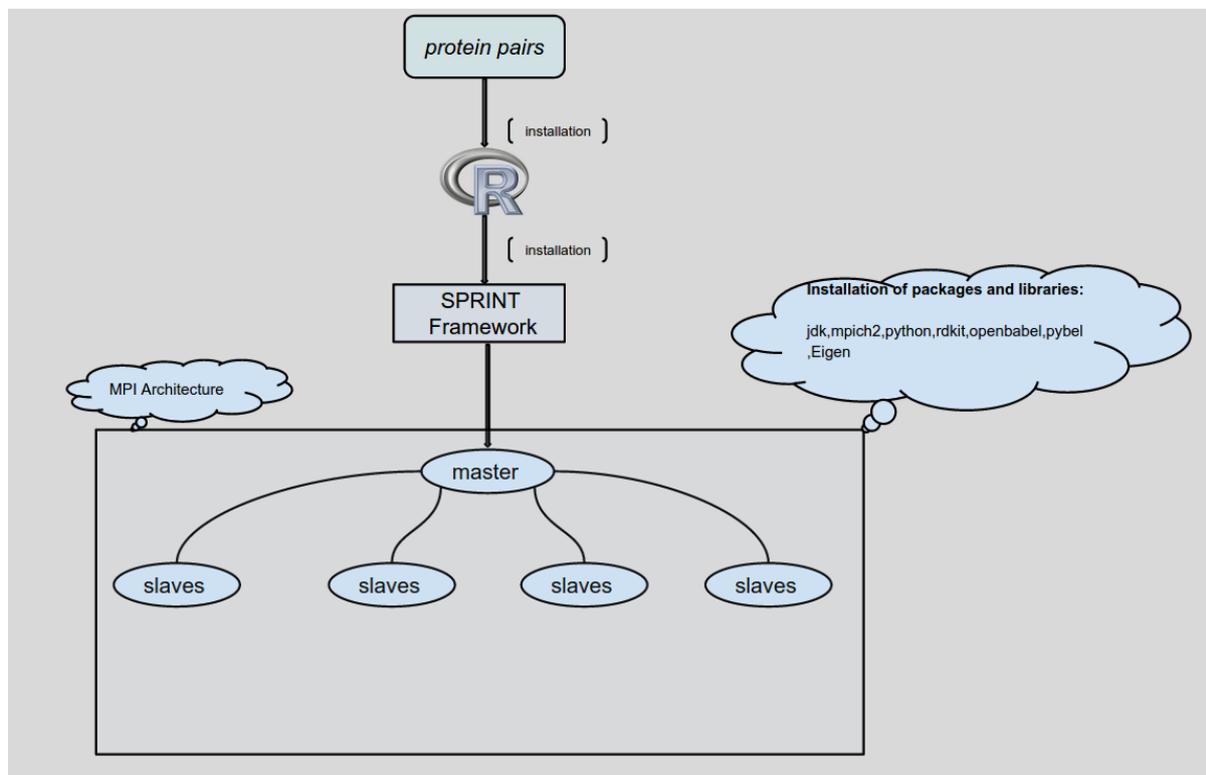

Figure 3.1: System Architecture

The details depicted in this figure are explained in what follows.

## 3.2 Pre-requisites

### 3.2.1 R language and SPRINT framework

Initially, we installed R (programming language) in the system. R is a programming language and software environment for statistical computing and graphics. The R language is widely used among statisticians and data miners for developing statistical software and data analysis. [1]

Subsequently, we installed the SPRINT framework. The original rationale for SPRINT is based on microarray analysis allowing the simultaneous measurement of thousands to millions of genes or sequences across tens to thousands of different samples. The analysis of the resulting data tests the limits of existing bioinformatics computing infrastructure. A solution to this issue is to use High Performance Computing (HPC) systems, which contain many processors and more memory than desktop computer systems. Although these form the original basis of SPRINT, most of existing parallelized functions are in principle useful in any situation where a very large number of computations are carried out or where computations results in very large memory use. This framework allows the addition of parallelized functions to R to enable the easy exploitation of HPC systems. The Simple Parallel R INTerface (SPRINT) is a wrapper around such parallelized functions. SPRINT allows R users to concentrate on the research problems rather than the computation, while still allowing exploitation of HPC systems. [2]

The following figure provides an overview of the SPRINT framework (3.2)

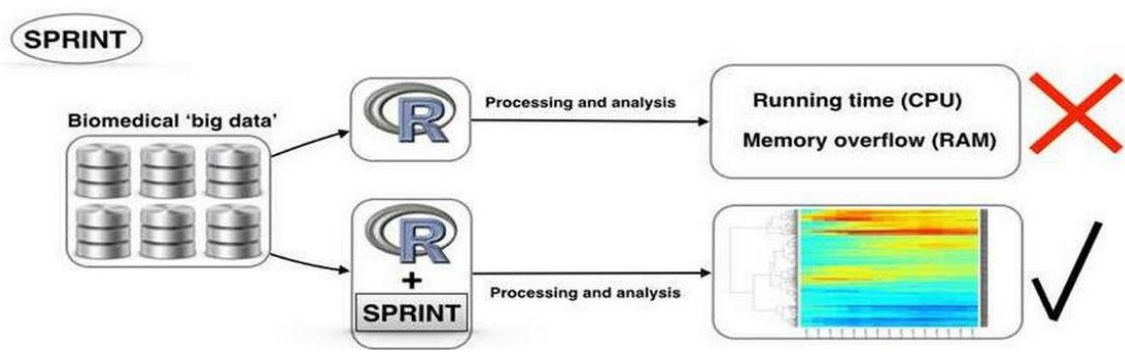

Figure 3.2: SPRINT framework (Source: http://www.ed.ac.uk/pathway-medicine/our-research/dpmgroups/ghazal-group/pathway-informatics/sprint/prototype)

As mentioned above SPRINT is a framework that makes parallel algorithms available to R users. It is designed to be relatively easy to extend and is made up of two components:

- R<->SPRINT interface
- the compute cluster itself

Communication is accomplished via files.

The following figure represents the R<->SPRINT interface (3.3):

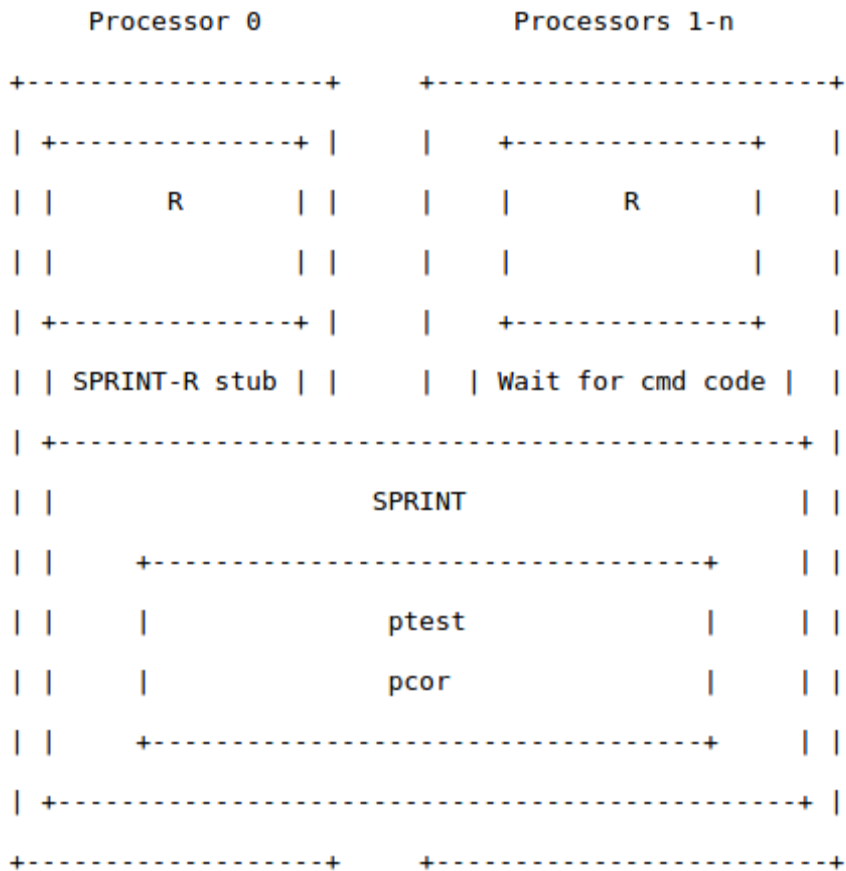

Figure 3.3: R<->SPRINT interface (Source: http://lists.r-forge.r-project.org/pipermail/sprint-developer/2013-August/000004.html)

When the user starts R with SPRINT loaded, the compute cluster goes into a wait state. When R reaches a function that is in SPRINT, the SPRINT-R stub sends a command to SPRINT via MPI. The MPI message contains an enumeration code that represents a function, which forces SPRINT to wake up and execute that function. The idea behind SPRINT is to allow parallel processing of data from within R without being restrained by R. Functions created in SPRINT should have similar interfaces to the serial R equivalent. Data required by the parallelized function is also passed via MPI. The creator of the function is responsible for creating that data flow. Afterwards, the function created is also responsible for passing the data back to R. This does not have to be the result of the processing, it could be a file handle, or a simple error code.

To create a new SPRINT function, it is required:

- creating an R stub (such that R can call the function)
- the C equivalent of the R stub
- the function to run in the computer cluster
- finally, connecting the different parts [3]

### 3.2.1.1 SPRINT Architecture

The SPRINT framework is made of two core components:

- An intelligent parallel harness that manages all access to the HPC resources hiding the complexity from the user.
- A flexible library of parallelized R functions that can be easily extended by adding more functions.

The parallelization model adopted is a task farm with a Master process controlling the execution of many Worker processes. All nodes are running R, executing the R script and loading the SPRINT library. When the SPRINT library has been loaded, the Master node takes charges of the execution. When the Master node encounters a parallel function, it distributes the work to be done amongst the available processes. The parallel harness uses C and MPI. The Master process coordinates the reading and writing to files which is performed simultaneously and in parallel by all worker processes using MPI/IO (Figure 3.4).

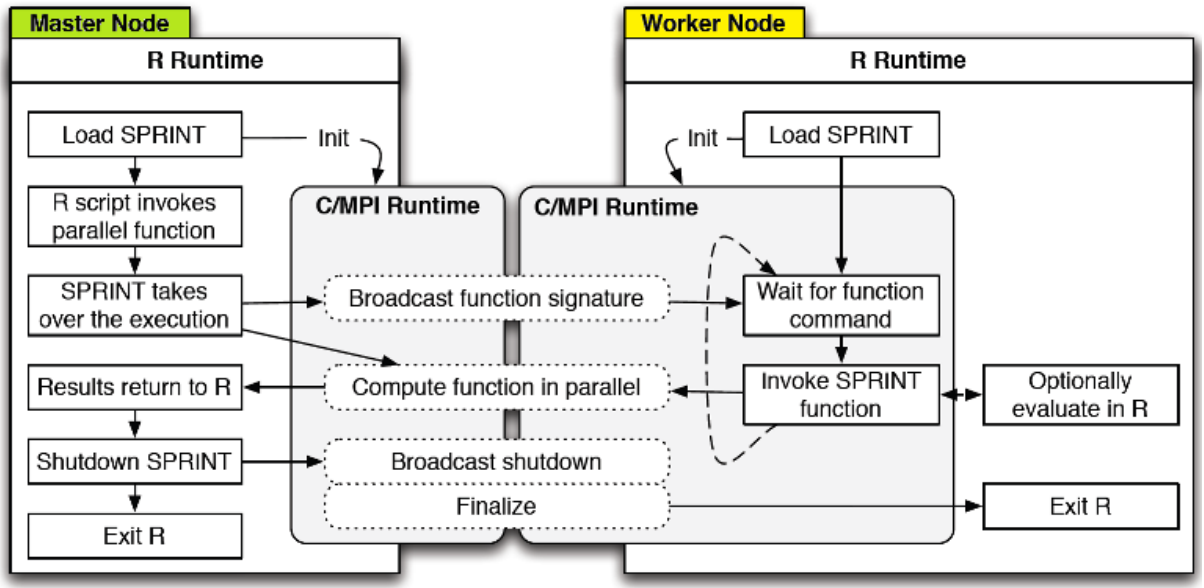

Figure 3.4: Architecture of SPRINT (Source: http://www.ed.ac.uk/pathway-medicine/our-research/dpmgroups/ghazal-group/pathway-informatics/sprint/sprint)

### 3.2.2 Message Passing Interface (MPI)

Message Passing Interface (MPI) is a standardized and portable message-passing system designed by a group of researchers from academia and industry to function on a wide variety of parallel computers. The standard defines the syntax and semantics of a core of library routines useful to a wide range of users writing portable message-passing programs in different computer programming languages such as Fortran, C, C++ and Java. There are several well-tested and efficient implementations of MPI, including some that are free or in the public domain. These fostered the development of a parallel software industry, and encouraged development of portable and scalable large-scale parallel applications. [6]

### 3.2.3 Python

Python is a widely used general-purpose, high-level programming language. Its design philosophy emphasizes code readability, and its syntax allows programmers to express concepts in fewer lines of code than would be possible in languages such as C++ or Java. The language provides constructs intended to enable clear programs on both a small and large scale. [7]

### 3.2.4 OpenBabel

Open Babel is a chemical toolbox designed to speak the many languages of chemical data. It's an open, collaborative project allowing anyone to search, convert, analyze, or store data from molecular modeling, chemistry, solid-state materials, biochemistry, or related areas. Also, Open Babel is a project to facilitate the interconversion of chemical data from one format to another – including file formats of various types. [8]

### 3.2.5 Pybel

Pybel provides convenience functions and classes that make it simpler to use the Open Babel libraries from Python, especially for file input/output and for accessing the attributes of atoms and molecules. The Atom and Molecule classes used by Pybel can be converted to and from the OBAtom and OBMol used by the openbabel module. [9]

### 3.2.6. RDKit

A collection of cheminformatics and machine-learning software is written in C++ and Python. The core algorithms and data structures are written in C++. Wrappers are provided to use the toolkit from either Python or Java. Additionally, the RDKit distribution includes a PostgreSQL-based cartridge that allows molecules to be stored in relational database and retrieved via substructure and similarity searches. [10]

# 4. Implementation

Inside SPRINT we implemented four (4) new functions that enable the orchestration of the pipeline in two levels of parallelization that allows the identification of the similarity between protein binding sites (BSS pipeline), either using the "broadcast" or "scatter-gather" MPI collective communication routines depending on the user's input:

- 2 functions for both levels of parallelization: MPI_Bcast
- 2 functions for both levels of parallelization: MPI_Scatter - MPI_Gather

To start the calculations and compare the two methods, we implemented R scripts inside SPRINT framework. Then, inside these scripts we created a matrix with proteins Id's and for all possible combinations of pairs, extracted specific results by the execution of Binding Site Similarity (BSS) pipeline. In this point, it is important to mention that the parallelization of the pipeline is conducted in two levels and for this reason we created two scripts for each function.

## 4.1 MPI_Bcast

A broadcast is one of the standard collective communication techniques. During a broadcast, one process sends the same data to all processes in a communicator. One of the main uses of broadcasting is to send out user input to a parallel program, or send out configuration parameters to all processes.

The communication pattern of a broadcast looks like this (Figure 4.1):

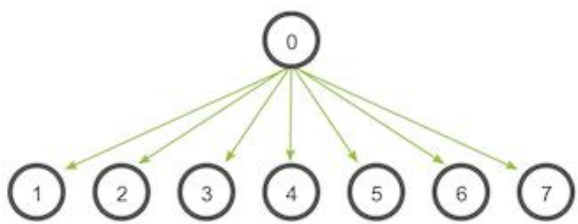

Figure 4.1: MPI_Bcast

In this example, process zero is the root process and it has the initial copy of data. All of the other processes receive the copy of data.

Although the root process and receiver processes do different jobs, they all call the same MPI_Bcast function. When the root process calls MPI_Bcast, the data variable will be sent to all other processes. When all of the receiver processes call MPI_Bcast, the data variable will be filled in with the data from the root process.

**4.1.1 MPI_Barrier**

One of the things that are important to mention about collective communication is that it implies asynchronization point among processes. This means that all processes must reach a point in their code before they begin executing again.

MPI has a special function that is dedicated to synchronizing processes:

*MPI_Barrier(MPI_Comm communicator)*

The name of the function is quite descriptive - the function forms a barrier, and no processes in the communicator can pass the barrier until all of them call the function. Below is presented an illustration in which the horizontal axis represents execution of the program and the circles represent different processes (Figure 4.2):

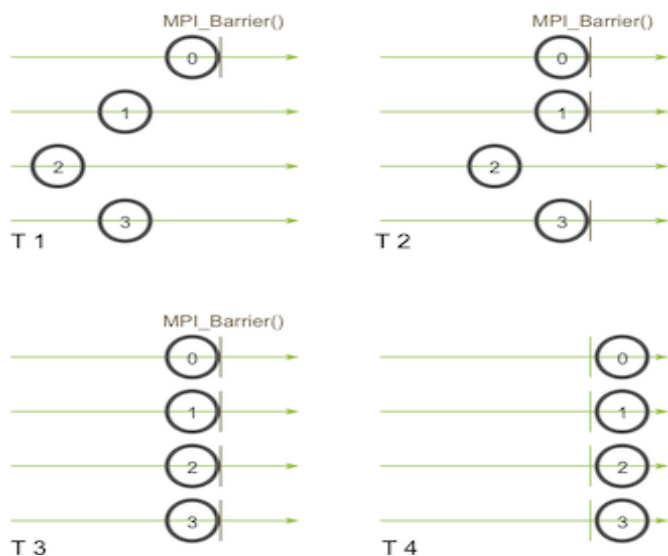

Figure 4.2: MPI_Barrier

Process zero first calls MPI_Barrier at the first time snapshot (T 1). While process zero is hung up at the barrier, process one and three eventually make it (T 2). When process two finally makes it to the barrier (T 3), all of the processes then begin execution again (T 4).

MPI_Barrier can be useful for many things. One of the primary use of MPI_Barrier is to synchronize a program so that portions of the parallel code can be timed accurately.

**4.2 MPI_Scatter - MPI_Gather**

As mentioned above we created also one more function to compare the computational performance between the two parallelization strategies.

**4.2.1 An introduction to MPI_Scatter**

MPI_Scatter is a collective routine that is very similar to MPI_Bcast. MPI_Scatter involves a designated root process sending data to all processes in a communicator. MPI_Bcast sends the same piece of data to all processes while MPI_Scatter sends chunks of an array to different processes.

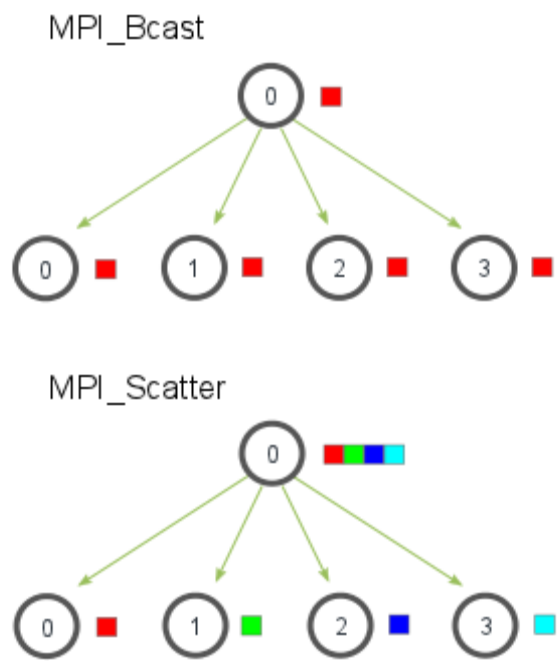

Figure 4.3: MPI_Bcast vs MPI_Scatter

In the illustration above (Figure 4.3), MPI_Bcast takes a single data element at the root process (the red box) and copies it to all other processes. MPI_Scatter takes an array of elements and distributes the elements in the order of process rank. The first element (in red) goes to process zero, the second element (in green) goes to process one, and so on. Although the root process (process zero) contains the entire array of data, MPI_Scatter will copy the appropriate element into the receiving buffer of the process.

**4.2.2 An introduction to MPI_Gather**

MPI_Gather is the inverse of MPI_Scatter. Instead of spreading elements from one process to many processes, MPI_Gather takes elements from many processes and gathers them to one single process. Below, is a simple illustration of this algorithm (Figure 4.4).

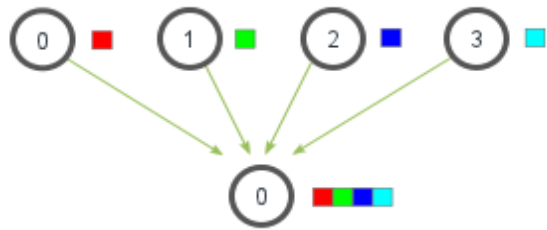

Figure 4.4: MPI_Gather

Similar to MPI_Scatter, MPI_Gather takes elements from each process and gathers them to the root process. The elements are ordered by the rank of the process from which they were received.

It is important to mention, that we used MPI_Barrier in the case of second function (MPI_Scatter - MPI_Gather).

**4.3 R scripts that call MPI_Bcast and MPI_Scatter-MPI_Gather in both levels of parallelization**

[Parallelization (broadcast)]()

> *R script(Level-1)*
>     *library("sprint")   // it loads SPRINT*
>     *psmapbc(matrix(c(pairs of protein Id's)))  // call function MPI_Bcast*
>     *pterminate()  // terminate R*
>     *quit() // quit R*
>
> *R script(Level-2)*
>     *library("sprint")   // it loads SPRINT*
>     *psmapbcV2(matrix(c(pairs of protein Id's)))  // call function MPI_Bcast*
>     *pterminate()  // terminate R*
>     *quit() // quit R*

## Parallelization (scatter-gather)

> ***R script(Level-1)***
> 
>     *library("sprint")   // it loads SPRINT*
> 
>     *psmapSG(matrix(c(pairs of protein Id's)))  // call function  MPI_Scatter-MPI_Gather*
> 
>     *pterminate()  // terminate R*
> 
>     *quit() // quit R*
> 
> ***R script(Level-2)***
> 
>     *library("sprint")   // it loads SPRINT*
> 
>     *psmapSGV2(matrix(c(pairs of protein Id's)))  // call function MPI_Scatter-MPI_Gather*
> 
>     *pterminate()  // terminate R*
> 
>     *quit() // quit R*

As we mentioned above, we implemented four new functions inside SPRINT framework called: ***psmapbc(), psmapbcV2()*** for both levels of parallelization which use broadcast and ***psmapSG()***, ***psmapSGV2()*** for both levels of parallelization which use scatter-gather.

In general, these functions enable the orchestration of the pipeline that allows the identification of the similarity between protein binding sites.

In practice, each of these functions receives an R matrix as input and implements the following steps:

a) creates all unique combinations (pairs of doubles) of the matrix' elements
b) distributes the pairs to the available processes (either using the "broadcast" or "scatter" MPI collective communication routines depending on the user's input)
c) runs a C code in each of the nodes using each pair of doubles as input

The interface and parameters to the function *psmapbc* are:

*psmapbc (x)*

where:

    x is an R matrix containing the set of elements.

Value x contains the unique protein IDs. The IDs are described in the Protein Data Bank repository [18]. Furthermore, in the third step of the algorithm each process executes the binding site similarity algorithm (BSS).

The interface and parameters described above are the same for the four functions and these have the role of coordinator of the entire process. The functions psmapbc() and psmapbcV2(), call smapbc() and smapbcV2() respectively that implement the third step and psmapSG(), psmapSV2() call smapSG() and smapSGV2() respectively that implement also the third step.

## 4.4 Binding Site Similarity (BSS) pipeline

The main plan is to conduct experiments based on the pipeline algorithms proposed by Haupt et al (2013) [11]. Below, it is presented an illustration of Binding Site Similarity (BSS) pipeline (Figure 4.5):

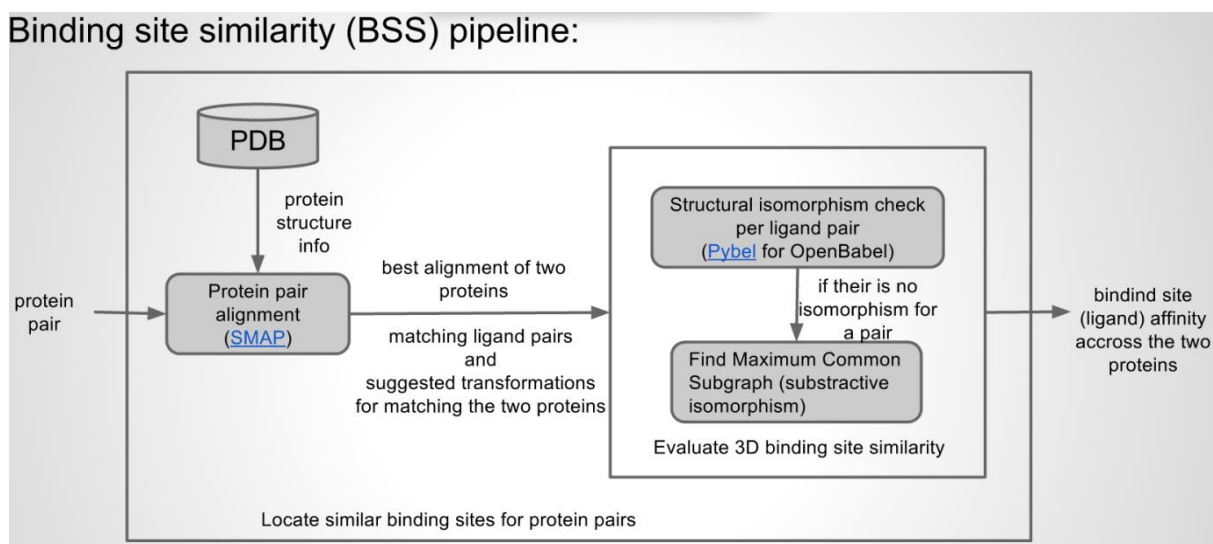

Figure 4.5: Binding Site Similarity (BSS) pipeline

BSS is accomplished in following steps:
- a protein pair is received as input
- for each protein, informations are extracted from PDB (3D shapes, nucleic acids, complex assemblies, etc.)
- SMAP is executed
    - ligands of each protein
    - best alignment of the two proteins
    - matching ligands pairs
    - transformations and rotations for matching the two proteins
- matching is conducted at the level of ligands
    - check structural isomorphism per ligand pair (use of Pybel and Openbabel)
    - subtractive isomorphism (maximum common subgraph if there is no isomorphism for the protein pair)
- ligand pairs of the two proteins are found and results for one pair with the minimum difference-maximum similarity are exported

The Protein Data Bank (PDB) is a repository for the three-dimensional structural data of large biological molecules, such as proteins and nucleic acids. The data, are typically obtained by X-ray crystallography or NMR spectroscopy and are submitted by biologists and biochemists from around the world, and are freely accessible on the Internet via the websites of its member organizations (PDBe, PDBj and RCSB). [4]

The original BSS code was written in Python with calls to Python library Pybel (and OpenBabel) and Java library SMAP. Binding Site Similarity (BSS) pipeline was originally designed to accept one protein pair and the execution conducted sequentially for specific ligands. Through our implementation, the execution accomplished for all the proteins is needed to be examined with all the possible ligand pairs and parallelized through our functions.

### 4.5 Optimizations

As we mentioned above we achieved to parallelize the BSS pipeline in two different levels. In first level, we parallelized the whole pipeline (protein alignment and ligand similarity) and in second level, we parallelized the slow parts of the pipeline (ligand matching).

### 4.5.1 First level of parallelization

We have wrapped the code for BSS pipeline (protein pair similarity) in SPRINT functions (blue box), that take as input a list of protein's pairs ids and examine each pair in a parallel thread. This is illustrated below (Figure 4.6):

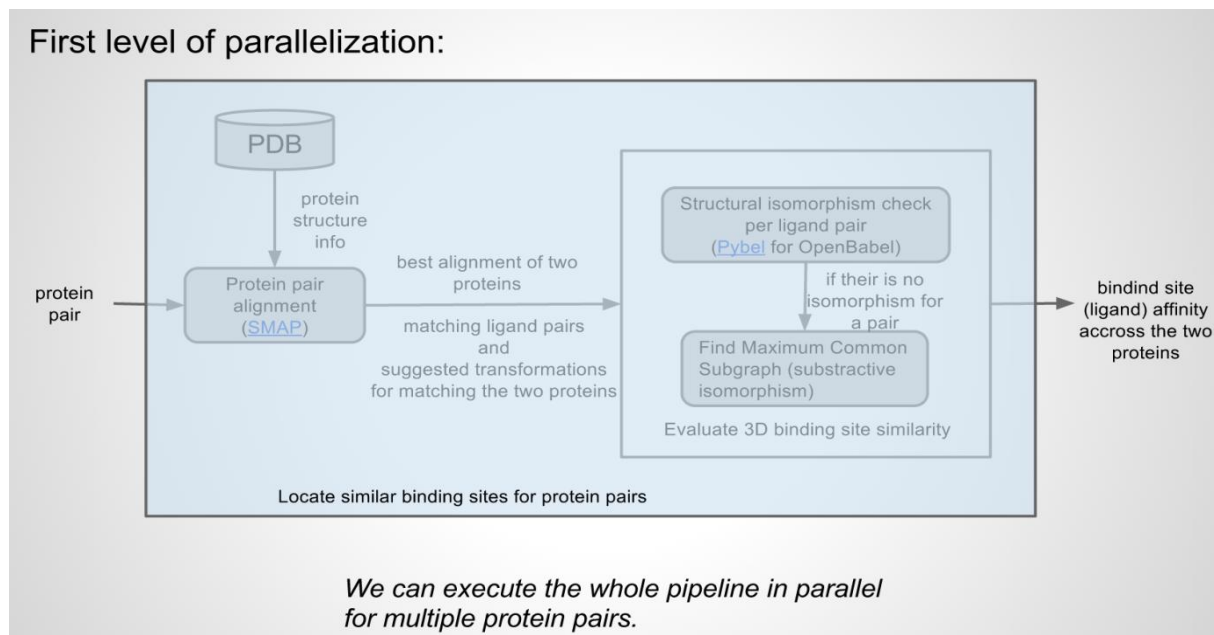

Figure 4.6: First level of parallelization

### 4.5.2 Second level of parallelization

We have parallelized the slow parts of BSS pipeline (ligand matching). We have modified the code for binding site similarity (green box) to examine each ligand pair in a separate thread. This is illustrated below (Figure 4.7):

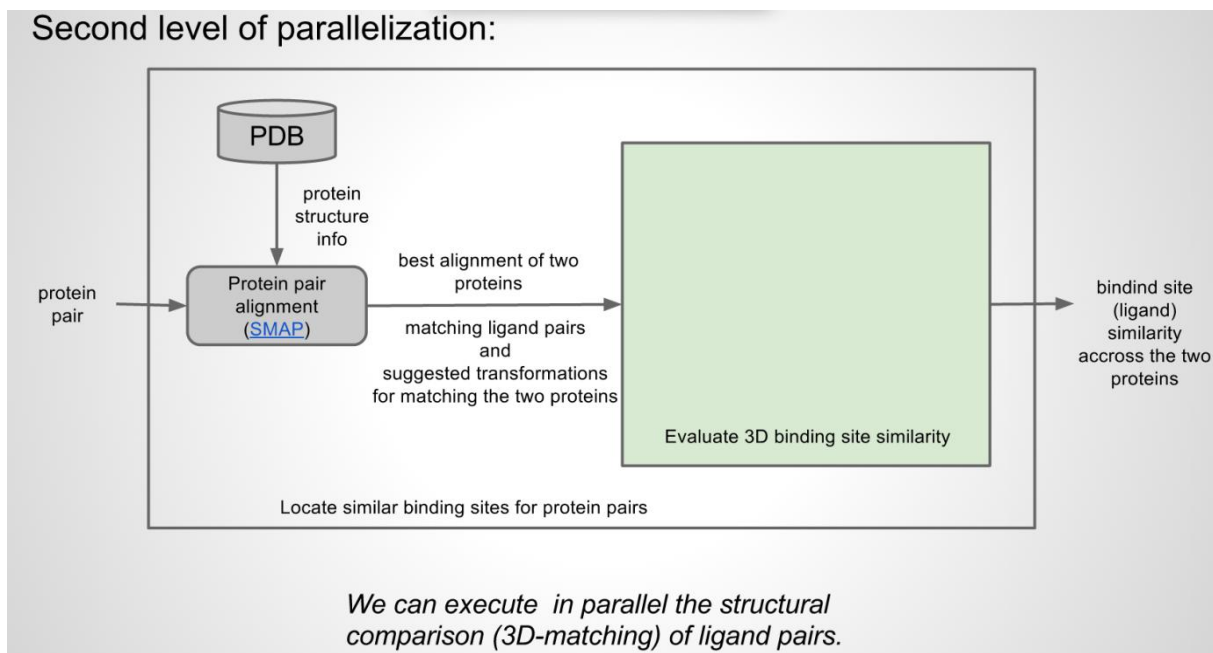

Figure 4.7: Second level of parallelization

For the need of this parallelization we use Python C-API. The Application Programmer's Interface to Python gives C and C++ programmers access to the Python interpreter at a variety of levels. The API is equally usable from C++, but for brevity it is generally referred to as the Python/C API. There are two fundamentally different reasons for using the Python/C API. The first reason is to write extension modules for specific purposes; these are C modules that extend the Python interpreter. The second reason is to use Python as a component in a larger application; this technique is generally referred to as embedding Python in an application. [5]

### 4.5.3 Implemented parallelization alternatives in SPRINT

We wrapped everything in SPRINT functions. The functions that we have created (MPI_Bcast, MPI_Scatter - MPI_Gather) execute each level of parallelization.

In Figure 4.8, are presented the two functions (psmapbc, psmapbcV2 -> broadcast) that parallelize the levels of BSS pipeline (psmabc for Level 1, psmapbcV2 for Level 2).

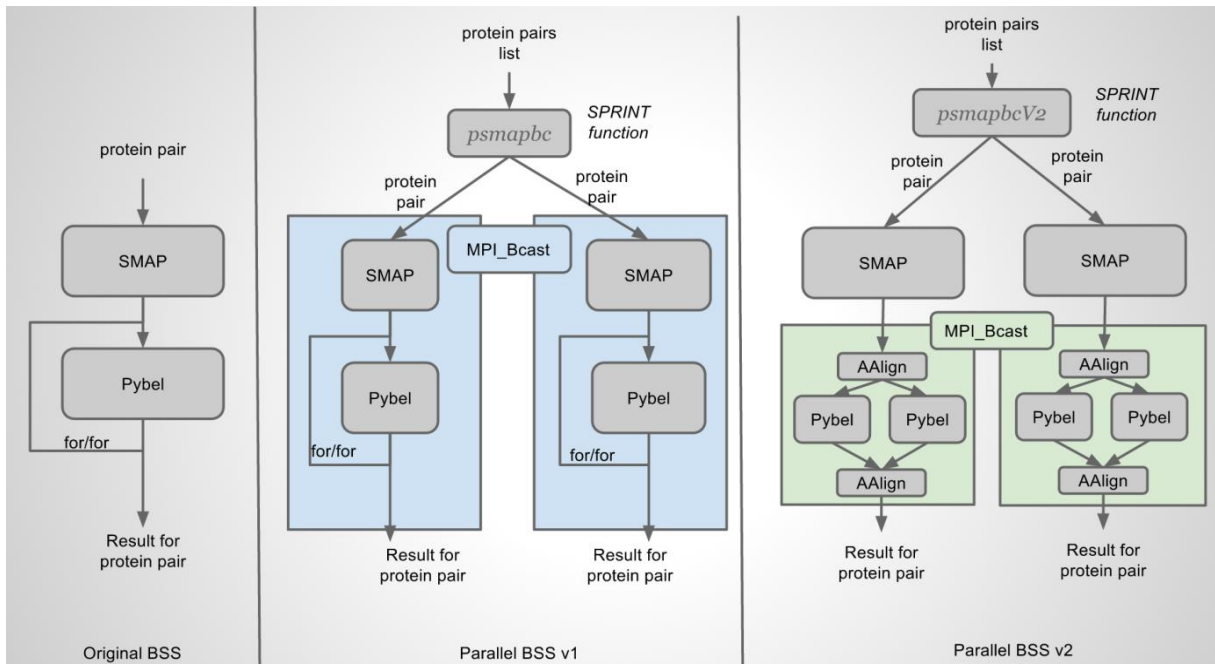

Figure 4.8: Functions (broadcast) that parallelize the two levels of BSS pipeline

In Figure 4.9, are represented the two functions (psmapSG, psmapSGV2 -> scatter-gather) that parallelize the levels of BSS pipeline (psmapSG for Level 1, psmapSGV2 for Level 2).

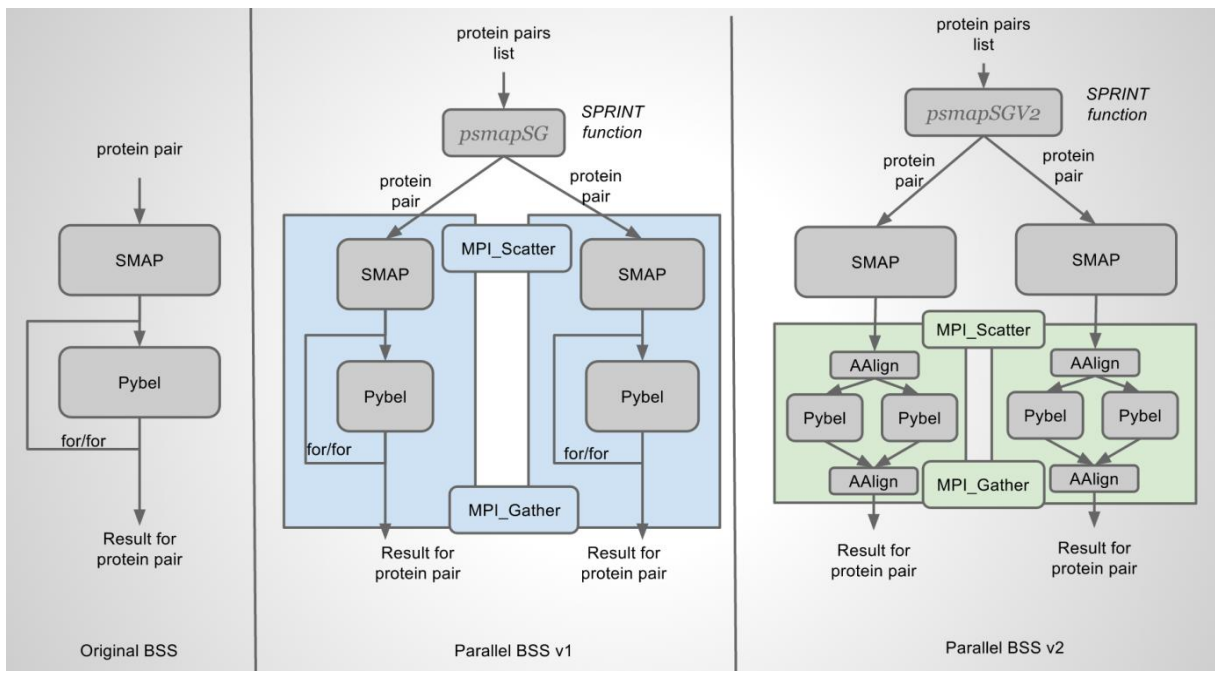

Figure 4.9: Functions (scatter-gather) that parallelize the two levels of BSS pipeline

# 5. Experiments - Results

## 5.1 Dataset description

For the purpose of experiments, we used six proteins ids (1iei, 1z89, 3p2v, 3kwb, 2bdl, 2auz) and according to mathematic formula *(n^2 - n)/2,* 15 unique combinations of protein pairs were finally extracted.

Figure 5.1 below represents the number of ligands combination per protein pairs.

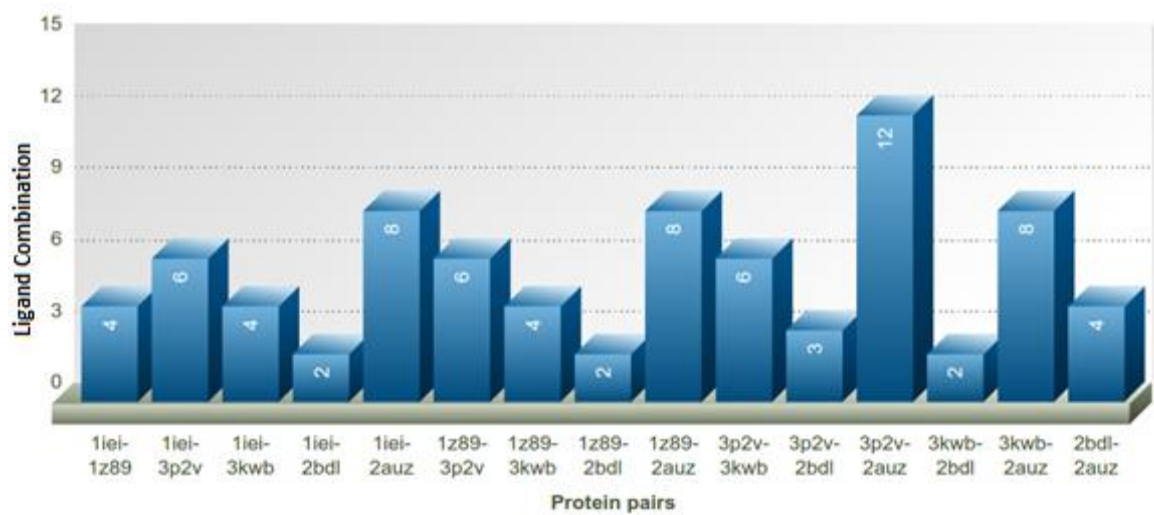

Figure 5.1: Number of ligands combination per protein pairs

Figure 5.2 represents the functions that used for ligand comparison.

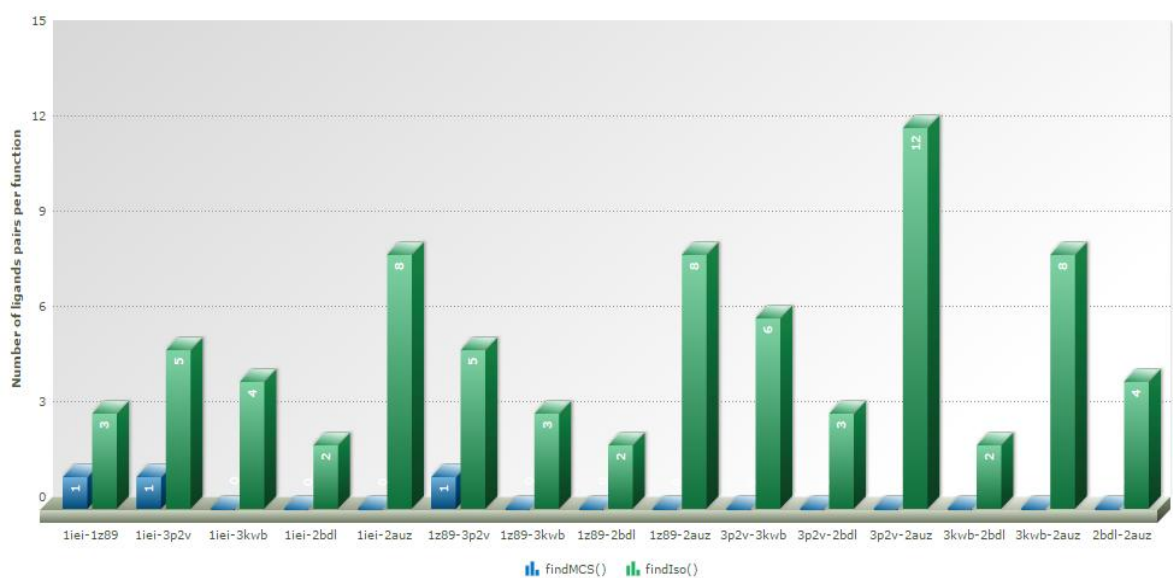

Figure 5.2: Isomorphism and MCS function calls per ligand pair

In the Binding Site Similarity code there are two functions that perform ligand comparison for the protein pairs. These are findIso() that finds all isomorphisms between the target and the query mol and findMCS() that performs substructure search if there is not isomorphism. In effect, this function computes the maximum common subgraph of the two molecules and finds substructure isomorphisms.

## 5.2 One node setup - Laptop vs okeanos 1 machine

In order to compare the results of our experiments, we execute the BSS pipeline (both levels of parallelization) in our machine but we also use okeanos infrastructure. In this infrastructure it is possible to create virtual machines and particularly constitutes a IaaS Service. "IaaS" stands for "Infrastructure as a Service", and it means that anyone can build virtual machines (computers), always connected to the Internet, without problems such as hardware failures, spaghetti cables, connectivity hiccups and software troubles.

### 5.2.1 Configuration of machines

Laptop configuration:
- OS: Ubuntu Desktop 14.04 LTS
- 4 GB RAM
- 2 CPUs

Okeanos 1 machine configuration:
- OS: Ubuntu Server 14.04 LTS
- 6 GB RAM
- 4 CPUs

### 5.2.2 Single Node – 1st level of parallelization

In this experiment, we measure the average time (5 runs) needed for the execution of the whole pipeline for 15 protein pairs. We implement the 1st level of parallelization in which each protein pair is processed in a separate thread using broadcast. The times reported in Table 1 are measured in the laptop using 4 threads.

Table 1. Results of single node – 1st level of parallelization experiment

| EXECUTIONS | TIME(seconds) |
|---|---|
| 1 | 225,392034 |

| | | |
|---|---:|---|
| | 2 | 224,697258 |
| | 3 | 225,952008 |
| | 4 | 229,972001 |
| | 5 | 228,372237 |
| **Average time** | | **226,8771076**±1.94 at 95% CI |

From the results of Table 1, we see that the worst execution time for the 15 protein pairs is a bit less than 4 minutes.

### 5.2.3 Single Node – 2nd level of parallelization

Table 2. Results of single node – 2nd level of parallelization experiment

| EXECUTIONS | TIME(seconds) |
|---:|---|
| 1 | 208,757258 |
| 2 | 208,045365 |
| 3 | 207,767564 |
| 4 | 209,978678 |
| 5 | 207,857208 |
| **Average time** | **208,48122**±0.81 at 95% CI |

From the results we can see that the 2nd level of parallelization offers a significant improvement in execution time, at the level of 8% of the initial time, with the same hardware setup.

### 5.2.4 Single Node Okeanos – 1st level of parallelization

In this experiment, we measure the average time (5 runs) needed for the execution of the whole pipeline for 15 protein pairs. We implement the 1st level of parallelization in which each protein pair is processed in a separate thread using broadcast. The times reported in Table 3 are measured in the okeanos infrastructure using 4 threads and 4 CPUs.

Table 3. Results of single node in okeanos – 1st level of parallelization experiment

| EXECUTIONS | TIME(seconds) |
|---:|---|
| 1 | 165,289306 |
| 2 | 165,603692 |
| 3 | 166,597197 |
| 4 | 164,852996 |
| 5 | 167,377057 |
| **Average time** | **165,9440496**±0.9 at 95% CI |

From the results we can see that the 1st level of parallelization in okeanos infrastructure offers a significant improvement in execution time, compared with laptop setup. This is because in okeanos, we have more resources and data spread over several CPUs.

### 5.2.5 Single Node Okeanos – 2nd level of parallelization

Table 4. Results of single node okeanos – 2nd level of parallelization experiment

| EXECUTIONS | TIME(seconds) |
|---:|---|
| 1 | 148,991805 |
| 2 | 147,961708 |
| 3 | 148,786597 |
| 4 | 147,564523 |
| 5 | 148,453654 |

| | Average time | 147,84±0.52 at 95% CI |
|---|---|---|

From the results we can see that the 2nd level of parallelization in okeanos, offers the best improvement in execution time. We observe that the execution time compared with laptop single node 2nd level of parallelization, offers a significant improvement in execution time, at the level of 28%.

### 5.2.6 Discussion of results

Figure 5.3 illustrates the time comparisons between Laptop - Okeanos setups in second level of parallelization.

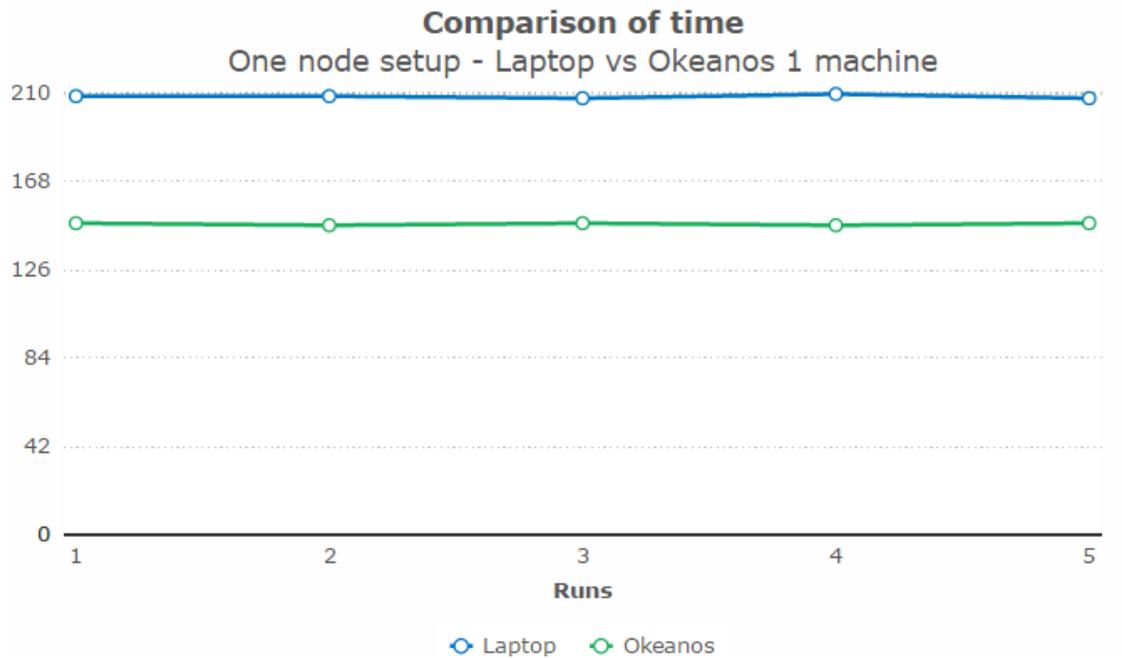
Figure 5.3: Time comparison in second level of parallelization

Figure 5.4 illustrates the average time comparisons between Laptop - Okeanos setups, in both levels of parallelization.

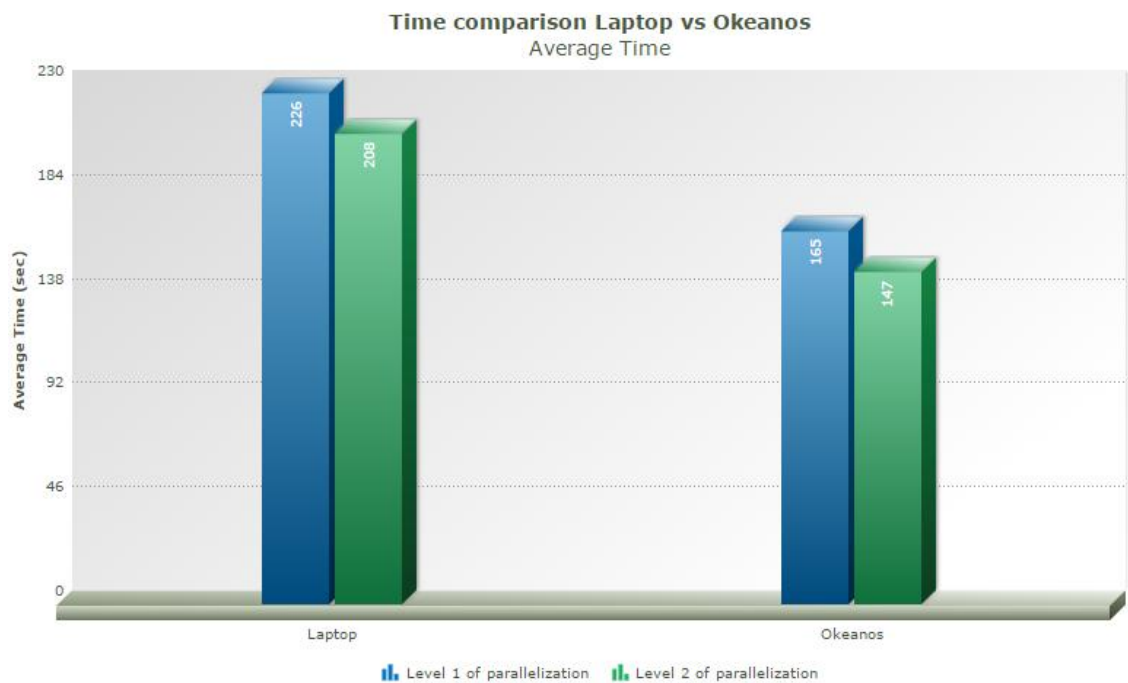

Figure 5.4: Average time comparisons in both levels of parallelization

As we see in the above figures the execution time between the two levels of parallelization is dropping significantly. This is because in the second level, we have parallelized the slow parts of BSS pipeline (ligand matching).

Also in okeanos setup we see important reduction of time execution in both parallelizations versus laptop setup, because the infrastructure is implemented with 4 CPUs.

Figure 5.5 illustrates the elapsed time of the whole process but also the time of findIso() and findMCS(). For this experiment, we used the second level of parallelization in okeanos infrastructure.

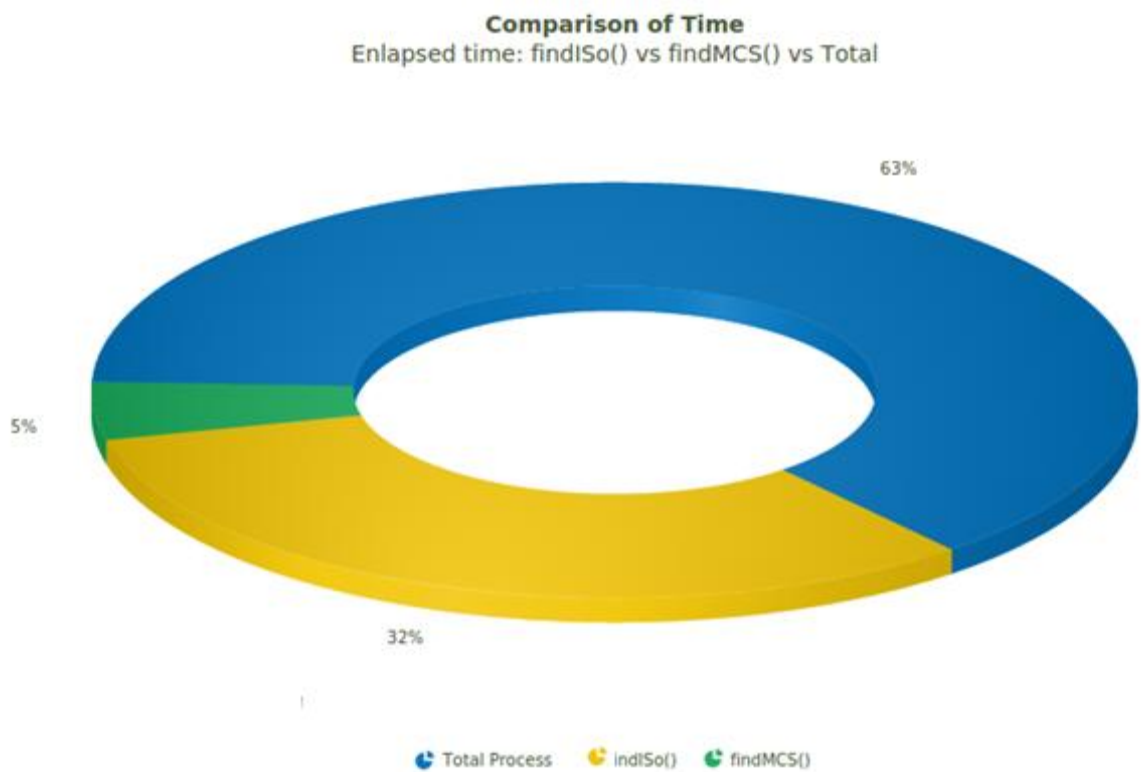

Figure 5.5: Elapsed time findIso() vs findMCS() vs Total_time of the remaining processing

Figure 5.6 illustrates the elapsed time of the whole process but also the time of findIso() and findMCS(), for one protein pair (1ei3,1z89). In this figure it is clear how much time is consumed for the execution of the functions for ligand comparison.

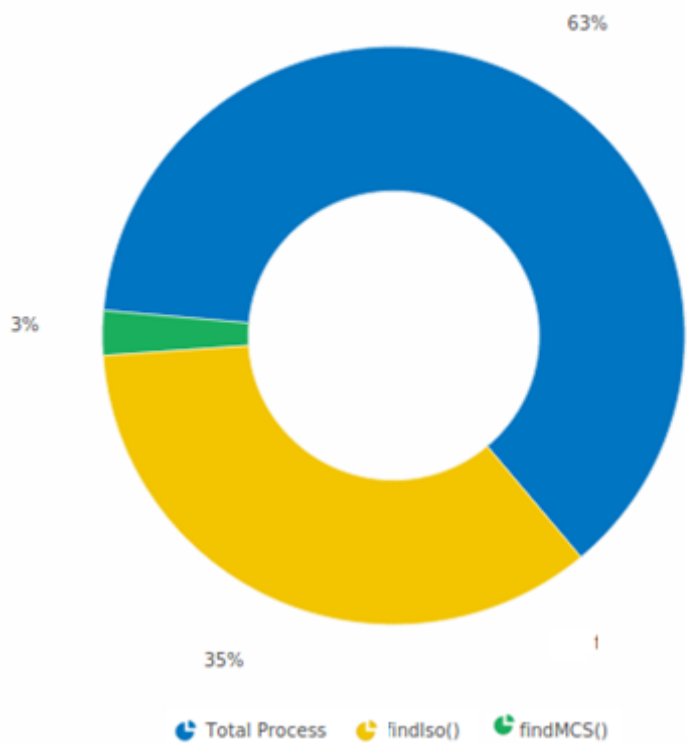

Figure 5.6: Elapsed time findIso() vs findMCS() vs Total time of the remaining processing for one protein pair (1iei,1z89)

The total runs of findMCS() and findIso(), used for ligand comparison is 79. The total runs of function findMCS() is *3* and the total runs of function findIso() is *76*.

From the above diagrams, we can see that 3 runs of the function findMCS() occupies 3% of the total time of the remaining processing which is 97,77 %, ie approximately 4,8 seconds. On the other hand, we observe that 76 runs of the function findIso() occupies 35% of the total time of the remaining processing, ie approximately 52 seconds.

# 6. Conclusion

As we mentioned above, drug repositioning process makes use of technologies and algorithms based on sequential programming.

Over the years there has been increasing data that they need treatment and study in the field of polypharmacology. This leads to the need for parallel programming technologies. As a result, we achieved to parallelize this process, namely the Binding Site Similarity pipeline.

Summarizing the results of this thesis, through our architecture and technology implementation, the accuracy and performance of the system was improved.

As noticed by the results of experiments, the execution time of both parallelization is reduced, when we ran our experiments on Okeanos using more cores. Ie the more resources used, the better runtime had. Also it was observed significant differences between the time between the two levels of parallelization. In first level, we parallelized the whole pipeline (protein alignment and ligand similarity) and in second level, we parallelized the slow parts of the pipeline (ligand matching). We noticed a reduction of around 50%, something which is a significant result in performance.

## 6.1 Future Work

As observed from the experiments carried out, there are points in our implementation that present bottlenecks. For example a large part of the total time, occupied downloading the proteins we want to examine from the PDB. One possible solution is to save the files to a local repository once before the implementation of BSS pipeline, and pull from there the data needed each time.

Also we aim to adapt to existing implementation, the ProBis algorithm. Unlike SMAP performed sequence alignment and exported among others translation and rotations for the proteins, PROBIS performs four times this process and selects the best case, ie does not take into consideration ligands that their distances are far ( according P-VALUE).

Furthermore as we have seen from our experiments a large part of the overall process occupies the functions findMCS() and findIso(). As mentioned above, at the level of ligand matching performed two functions. FindIso() that finds all isomorphisms between the target and the query mol and findMCS() that performs substructure search if there is not isomorphism. In effect, this function computes the maximum common subgraph of the two molecules and finds substructure isomorphisms. There would be important to see, what information can be cached regarding the molecules contemplated, so that upon the execution of these functions be not extracted information all over again for the same ligands. In this way, we believe that the execution time will improve.

Moreover, within the next few months we aim to implement one more function in the SPRINT framework for clustering purposes. The method will implement a parallelized version of the popular hierarchical agglomerative clustering (HAC) algorithm. In comparison to partitioning algorithms (e.g. k-medoids, PAM), HAC produces a hierarchy of clustering schemes as output.
The implementation will be based on PINK, a distributed-memory parallel algorithm through solving the minimum spanning tree problem with Prim's algorithm. PINK, is described in "A Scalable Algorithm for Single-Linkage Hierarchical Clustering on Distributed-Memory Architectures". [17] Among others, we plan to compare the results of ppam against the HAC implementation on SPRINT.